%% file: ChiralLetter.tex
\documentclass[DIV15,12pt,a4paper,headinclude]{scrartcl}
\usepackage[english]{babel}
\usepackage{amsmath}
\usepackage{amssymb}
\usepackage{bbm}
\usepackage{empheq}
\usepackage{mathrsfs}
\usepackage{mathtools}
\usepackage{array}
\usepackage{dsfont}
\usepackage{graphicx}
\usepackage{psfrag}
\usepackage{pstricks}
\usepackage{color}
\usepackage{booktabs}
\usepackage[normal,margin=10pt,font=small,labelfont=bf,
labelsep=period]{caption}
\usepackage[hang]{subfigure}
\usepackage[onehalfspacing]{setspace}
\usepackage{placeins}
\usepackage[bottom]{footmisc} 
\usepackage{indentfirst}
\usepackage{overpic}

\numberwithin{equation}{section}

\newcommand{\Ofourloc}{\ensuremath{{\sf O}(4)_{\rm loc}}}
\newcommand{\Ofour}{\ensuremath{{\sf O}(4)}}
\newcommand{\wrt}{w.\,r.\,t.\ }
\newcommand{\ie}{i.\,e. }

\newcommand{\oab}{\ensuremath{\omega^{ab}_{\ \ \mu}}}

\newcommand{\bD}{\beta_{\rm D}}
\newcommand{\aD}{\alpha_{\rm D}}

\let\OLDthebibliography\thebibliography
\renewcommand\thebibliography[1]{
  \OLDthebibliography{#1}
  \setlength{\parskip}{0pt}
  \setlength{\itemsep}{0pt plus 0.3ex}
}

\begin{document}
\renewcommand{\baselinestretch}{1.1}
\begin{titlepage}
\enlargethispage{2\baselineskip}
\title{
\vspace{-1.5cm}
\begin{flushright}
\rm
\normalsize{MITP/15-076}
\bigskip
\vspace{0.5cm}
\end{flushright}
\rm On selfdual spin-connections and \\Asymptotic Safety}
\date{}
\author{{\large U. Harst\footnote{\ Email address: harst@thep.physik.uni-mainz.de} \ and M. Reuter\footnote{\ Email address: reuter@thep.physik.uni-mainz.de}}\\
{\small Institute of Physics, University of Mainz}\\[-0.4cm]
{\small Staudingerweg 7, D-55099 Mainz, Germany}}
\maketitle
\thispagestyle{empty}
\begin{abstract} 
\vspace{-1.5cm}
\renewcommand{\baselinestretch}{1.0}\selectfont
\noindent We explore Euclidean quantum gravity using the tetrad field together with a selfdual or anti-selfdual spin-connection as the basic field variables. Setting up a functional renormalization group (RG) equation of a new type which is particularly suitable for the corresponding theory space we determine the non-perturbative RG flow within a two-parameter truncation suggested by the Holst action. We find that the (anti-)selfdual theory is  likely to be asymptotically safe. The existing evidence for its non-perturbative renormalizability is comparable to that of Einstein-Cartan gravity without the selfduality condition. \end{abstract}
\end{titlepage}
\newpage 

\section{Introduction}
While it has been clear for several decades that besides the metric approach to General Relativity there exists an essentially equivalent description in terms of tetrads and spin-connections (``Einstein-Cartan gravity''), one of the important surprises uncovered by the work of Ashtekar \cite{Ash-Ham, A1, A2, martinrev} was that after a canonical transformation to new field variables the spin-connection may be chosen {\it selfdual} (or anti-selfdual). For Lorentzian signature and structure group ${\sf O}(1,3)$, selfdual connections are unavoidably complex, which complicates their quantization. In the Euclidean case, they are real, however, and the condition of selfduality precisely halves the number of the connection's independent (real) components. In the generic case when the connection is not necessarily (anti-)selfdual, the Euclidean form of Ashtekar's theory can be obtained from the Holst action \cite{Holst1996}; it depends on the tetrad $e_a^{\ \mu}$ and on the spin-connection $\omega^{ab}_{\ \ \mu}$ via its curvature $F^{ab}_{\ \ \mu\nu}$:

\begin{equation}
S_{\rm Ho} [e, \omega]= -\frac{1}{16 \pi G} \int {\rm d}^4x\, e \bigg[ e_a^{\ \mu} e_b^{\ \nu} \bigg(F^{ab}_{\ \ \mu\nu} - \frac{1}{2 \gamma} \varepsilon^{ab}_{\ \ cd}F^{cd}_{\ \ \mu\nu}\bigg)-2\Lambda\bigg]
\end{equation}
 
\noindent Besides the two terms known from the familiar first-order approach to general relativity, $S_{\rm Ho}$ contains a third one, containing the a priori arbitrary Immirzi parameter, $\gamma$, \cite{Immirzi:1996di, Barbero}.

The case of $\gamma=\pm 1$ is special as the action $S_{\rm Ho}$ then only depends on one chirality of the spin connection \oab, \ie on its selfdual or anti-selfdual part \wrt the \Ofour-indices. The (Euclidean!) duality operator being defined as $(\star)^{ab}_{\ \ cd}= \frac{1}{2} \varepsilon^{ab}_{\ \ cd}$ squares to unity, and objects with eigenvalue $+1$ are called selfdual, those with eigenvalue $-1$ anti-selfdual. We can define a projector on the (anti-)selfdual part of any antisymmetric second rank \Ofour-tensor by $(P^{\pm})^{ab}_{\ \ cd}=\frac{1}{4}(\delta^a_{[c}\delta^b_{d]}\pm\varepsilon^{ab}_{\ \ cd})$ and decompose it into a selfdual and anti-selfdual part, e.g. $F^{ab}=F^{(+)\,ab}+F^{(-)\,ab}$. In $S_{\rm Ho}$ the combination of curvature and Immirzi term in the case of $\gamma=\pm 1$ leaves us with
\begin{equation}
 \frac{1}{2}\!\int\!\! {\rm d}^4x \,e\!\!\left[e_a^{\ \mu} e_b^{\ \nu}\! \left(\!F^{ab}_{\ \ \mu\nu}\!\mp \frac{1}{2} \varepsilon^{ab}_{\ \ cd}F^{cd}_{\ \ \mu\nu}\!\right)\right]=\int\!\! {\rm d}^4x\, e\!\left[e_a^{\ \mu} e_b^{\ \nu} \left(P^{\mp\,ab}_{\ \ \ \ cd}F^{cd}_{\ \ \mu\nu}\right)\right].
\end{equation}

\noindent One can show that the (anti-)selfdual part of the field strength tensor of a generic spin connection equals the field strength tensor of the (anti-)selfdual part of that spin connection: $(P^{\pm}F)^{ab}(\omega)=F^{(\pm)\,ab}(\omega)=F^{ab}(\omega^{(\pm)})=F^{ab}(P^{\pm}\omega)$. As a result, there are only 12 (rather than the usual 24) independent components of the spin connection $\omega^{(+)}$ or $\omega^{(-)}$ on which the action really depends when $\gamma=\mp 1$. Thus $S_{\rm Ho}\equiv S[e, \omega^{\pm}]$ for these special values of the Immirzi parameter, and this action leads to 12 equations of motion, that can be solved for $\omega^{(\pm)}(e)$, when the invertibility of the tetrad is assumed \cite{Giulini:AshtekarVar}. One can show that $\omega^{(\pm)}(e)$ is the (anti-)selfdual projection of the spin connection corresponding to the Levi-Civita connection, $\omega^{(\pm)}(e){}^{ab}{}_{\mu}=(P^\pm \omega_{\rm LC})^{ab}{}_\mu$. This spin connection necessarily gives rise to a spacetime with a non-vanishing torsion \cite{Hehl-2, Shapiro2002}. Nonetheless, substituted into the tetrad equations of motions we, again, arrive at Einstein's equation. Thus we find their solutions among the classical solutions of selfdual gravity, albeit formulated in a spacetime with a connection differing from the usual Levi-Civita connection. In classical gravity without fermionic matter \cite{Freidel:Immirzi1,Perez2003,Perez_Rovelli:Immirzi_Parameter} this difference cannot be observed \cite{Gi-Li-1, Gi-Li-2}. 

In quantum-dynamical computations, in particular at the off-shell level, differences can, and do occur, however. For the case where non-selfdual connections and tetrads were chosen to serve as the basic field variables these differences were studied in refs. \cite{je:proc,je:lett,je:longpaper,long} by means of a functional renormalization group equation (FRGE). Hereby the main emphasis was on the possibility that the theory might be nonperturbatively renormalizable along the lines of the Asymptotic Safety scenario \cite{wein, mr, livrev, frankrev, frank1, perper1a, oliver2}. A perturbative investigation was reported in \cite{bene-speziale,bene-speziale2}. Given the large number of theories classically equivalent to, or observationally indistinguishable from General Relativity \cite{Peldan, A1, Pleb,CDJ1,CDJ2, Krasnov1, Krasnov2, Mielke-newvar,Nieh-Yan,Zan-Chan} it is conceivable that there exist several inequivalent, asymptotically safe quantum gravity theories \cite{Harst2012,je:lett,gregor1}.

For the case $\gamma=\pm 1$, to be studied in the present paper, the Holst action comprises a theory of gravity in (anti-)\allowbreak selfdual variables that depends on less independent field components. Therefore, when we try to compute a path integral over this action for a general value of $\gamma$, we have to expect divergences in the limit $\gamma\rightarrow \pm 1$, as the integration over the other duality component will not be suppressed at all. In order to set up a FRGE for the (anti-)selfdual case we thus have to eliminate the redundant field components before the operator traces on the RHS of the flow equation are evaluated. It will turn out that this elimination is rather straightforward if we employ the particular decomposition of the fluctuation fields advocated in \cite{long}. This way we are able to study in this paper the nonperturbative RG flow of gravity in selfdual variables for the first time.

\section{Nonperturbative RG flow of selfdual gravity}\label{QECG:Chiral}
The starting point of the present investigation is the FRGE-based analysis of (non-selfdual) Einstein-Cartan quantum gravity that was performed in ref. \cite{long}. In this analysis the effective average action was approximated by $\Gamma_k= \Gamma_{{\rm Ho}\,k}+ \Gamma^{\rm gf}_k+\Gamma^{\rm gh}_k$. Here $\Gamma_{{\rm Ho}\,k}$ denotes the Holst action $S_{\rm Ho}$ with running parameters $(G_k, \Lambda_k, \gamma_k)$, while $\Gamma^{\rm gf}_k$ and $\Gamma^{\rm gh}_k$ are the gauge-fixing and Faddeev-Popov ghost terms corresponding to the diffeomorphism and ${\sf O}(4)_{\rm loc}$ gauge conditions ${\cal F}_\mu=\frac{1}{\sqrt{\alpha_{\rm D}}}\bar{e}_a^{\ \nu}(\bar{\cal D}_\nu \varepsilon^a_{\ \mu} + \beta_{\rm D}\bar{\cal D}_\mu \varepsilon^a_{\ \nu})$ and ${\cal G}^{ab}= \frac{1}{\sqrt{\alpha_{\rm D}}} \bar{g}^{\mu\nu} (\bar{e}^b_{\ \nu} \varepsilon^{a}_{\ \mu}-\bar{e}^a_{\ \nu} \varepsilon^{b}_{\ \mu})$ respectively. They contain three $k$-independent gauge parameters $(\alpha_{\rm D}, \alpha_{\rm L}, \beta_{\rm D})$. Using the same notation and conventions as in \cite{long}, $\varepsilon^a_{\ \mu}\equiv e^a_{\ \mu}- \bar{e}^a_{\ \mu}$ and $\tau^{ab}_{\ \ \mu}\equiv \omega^{ab}_{\ \ \mu}- \bar{\omega}^{ab}_{\ \ \mu}$ denote the fluctuations of the tetrad and the spin-connection, respectively, and $\bar{\cal D}_\mu$ is the covariant derivative which contains both the (background) spacetime- and spin-connection.

Our functional RG analysis of selfdual gravity will be carried out using the same Wegner-Houghton-like flow equation and adapted plane wave-based projection techniques as in \cite{long}, namely $\partial_t \Gamma_k = \frac{1}{2} \, D_t {\rm STr\,}\Big|_k {\rm ln}\,\big(\Gamma_k^{(2)}\big)$. Here ${\rm STr\,}\Big|_k $ indicated an IR regularization of the supertrace by a sharp cutoff of the momentum integral, and the derivative $D_t$ acts only on the explicit $t\equiv {\rm ln}\,(k)$-dependence due to this cutoff.

In the following we will only highlight the structural differences of the RG equations for selfdual gravity compared to Quantum Einstein-Cartan Gravity (QECG), in subsection \ref{Special_Features}, before we derive its $\beta$-functions in subsection \ref{Beta_Functions}, and proceed with the presentation of the resulting RG flow in subsection \ref{RG_Flow}.

\subsection{Special features of the selfdual case}\label{Special_Features}

{\bf \noindent Field content.}
The most obvious modification in comparison to QECG is that we restrict the field space of spin connections to one chirality. Since the projectors $P^{\pm}=\frac{1}{2}(\mathds{1}\pm \star)$ decompose any connection according to $\omega=(P^++P^-)\omega=\omega^{(+)}+\omega^{(-)}$, this restriction corresponds to halving its number of independent components. Thus we are left with $28=16+12$ independent field components of vielbein and spin connection, respectively. This is reflected in the dimension of the Hessian $\Gamma_k^{(2)}$ that in the (anti-)selfdual case corresponds to a $28\times 28$-matrix differential operator. We will see in a moment how an adapted decomposition of the fields gives rise to a simple reduction of the $40 \times 40$ Hessian of the general Holst truncation to the (anti-)selfdual case.

{\bf \noindent Gauge symmetry.} If we denote the six generators of the full ${\sf O}(4)$-gauge group by $M_{ab}$, with $M_{ab}=-M_{ba}$, by definition they satisfy the algebra
\begin{equation}
 [M_{ab},M_{cd}]=i (\delta_{ac} M_{bd}+\delta_{bd} M_{ac}-\delta_{bc} M_{ad}-\delta_{ad} M_{bc})\:.
\end{equation}
A simple computation reveals that the 3 generators $M^\pm_{ab}=(P^{\pm}M)_{ab}$ of each sign satisfy an algebra of the same form, individually, and that the generators of different $\star$-eigenvalue commute with each other, $[M^\pm_{ab},M^\mp_{cd}]=0$. Using the t'Hooft $\eta$-symbols \cite{tHooft:EtaSymbol} that map (anti-)\allowbreak selfdual \Ofour-tensors onto {\sf SO}(3)-vectors it is in fact easy to show that the generators $L^\pm_i=\frac{1}{4}\eta_{i}{}^{ab}M^\pm_{ab}$ satisfy the usual angular momentum algebra $[L_i,L_j]= i \varepsilon_{ij}{}^{k} L_k$. Thus the \Ofour-algebra decomposes into two chiral factors such that locally also the groups satisfy
\begin{equation}
 {\sf O}(4)\mathop{\widetilde{=}}{\sf SO}^{+}(3)\times {\sf SO}^{-}(3).
\end{equation}
This is the Euclidean counterpart of the decomposed Lorentz group {\sf SO}(3,1), which is well known to comprise two chiral {\sf SU}(2) components, too. But in contrast to our case the boost and rotation generators in {\sf SO}(3,1) obtain as {\it complex} linear combinations of the {\sf SU}(2) components. Moreover, there, the eigenvalues of $\star$ are $\mp i$, whence the (anti-)selfdual components of a real tensor are necessarily complex.

When we restrict ourselves to spin connections of one chirality we, thus, also reduce the gauge group to one chiral component of the above decomposition. In summary, we therefore conclude that the theory space of (anti-)selfdual gravity is reduced in both, the field content and the total symmetry group {\bf G}, and is hence given by the set of action functionals
\begin{equation}
 {\cal T}^\pm_{\rm EC}= \Big\{ A[e^a{}_{\mu},\omega^{\pm}{}^{ab}{}_{\mu},\cdots]\,\Big|\, \text{inv. under } {\bf G}={\sf Diff}({\cal M)} \ltimes {\sf SO}^\pm(3)_{\rm loc}\Big\}\:.
\end{equation}
Here the dots stand for additional background- and ghost-fields.

{\bf \noindent Gauge conditions and ghost fields.}
With the reduced gauge group at hand also the 6 gauge fixing conditions ${\cal G}_{ab}$ of the former \Ofourloc-group have to be reduced to only 3 that are needed to gauge-fix the remaining ${\sf SO}^\pm(3)_{\rm loc}$ component. Most easily this is done by a projection of ${\cal G}_{ab}$ to its (anti-)selfdual part, using now
\begin{equation}\label{ChiralGF}
 {\cal G}^\pm_{ab}=(P^\pm {\cal G})_{ab}\:.
\end{equation}
We apply the Faddeev-Popov procedure exactly as before, and find that in $S_{\rm gh}$ simply the \Ofour-ghost fields $\bar{\Upsilon}_{ab},\Upsilon_{ab}$ get replaced by their (anti-)selfdual components $\bar{\Upsilon}^\pm_{ab},\Upsilon^\pm_{ab}$. The diffeomorphism gauge-condition ${\cal F}_\mu$ gets modified only slightly, since the covariant derivative inside it now is constructed from the (anti-)selfdual spin connection.

{\bf \noindent Irreducible field parameterization.} In order to partially diagonalize the Hessian of the Holst action the fields representing small fluctuations about the background $(\bar e, \bar \omega)$ were parameterized by component fields that transform irreducibly. For the spin connection the corresponding decomposition of $\tau^{ab}_{\ \ \mu}\equiv \omega^{ab}_{\ \ \mu}-\bar{\omega}^{ab}_{\ \ \mu}$ reads
\begin{equation}\label{usual_decomp}
\tau^{ab}_{~\mu} (x) \!=\! \frac{\bar{\mu}^{ \frac{1}{2}}}{\sqrt{2}}\!\left[\!\frac{\partial_\mu \partial^{[a}}{ - \Box}A^{b]} (x)\!+\! \frac{\partial^{[a}}{\sqrt{\! - \Box}} B^{b]}_{~~\mu} (x)\! +\! \varepsilon^{ab}_{~\:cd}\frac{\partial_\mu \partial^c}{ - \Box} C^d (x)\! +\! \varepsilon^{ab}_{~\:cd} \frac{\partial^c}{\sqrt{\! - \Box}} D^d_{~\mu} (x)\!\right]
\end{equation}
All component fields $(A, B, C, D)$ are fully transverse; they vanish upon contraction with $\partial_a$ or $\partial_\mu$. Thanks to a judiciously chosen set of conventions employed in \cite{long}, the two sets of fields $(A, B)$ and $(C, D)$, respectively, switch their roles under dualization: $(A, B) \stackrel{\star}{\longleftrightarrow}(C,D)$. By introducing the new fields $A_\pm\equiv (A\pm C)/\sqrt{2}$ and $B_\pm\equiv (B\pm D)/\sqrt{2}$ we may therefore rewrite \eqref{usual_decomp} as
\begin{equation}\label{selfdual_decomp}
\tau^{ab}_{~\mu} (x) = 2\sum_\pm \left(\big(P^{\pm}\big)^{ab}_{\ \ cd} \frac{\partial_\mu \partial^c}{-\Box}A_\pm^d+\big(P^{\pm}\big)^{ab}_{\ \ cd} \frac{ \partial^c}{\sqrt{-\Box}} B^{\ d}_{\pm\ \mu}\right).
\end{equation}
We observe that now the fields $(A_+, B_+)$ and $(A_-, B_-)$ describe the selfdual and anti-selfdual components of the fluctuation field, respectively.

Up to this point both decompositions, \eqref{usual_decomp} and \eqref{selfdual_decomp}, are completely equivalent, and the RG flow of non-selfdual Einstein-Cartan gravity with a running Immirzi parameter can be obtained using either decomposition. Upon expanding the action to second order in the fluctuations this results in a decomposed quadratic form $\Gamma^{\rm quad}_{+}+\Gamma^{\rm quad}_{-}$ whereby $\Gamma^{\rm quad}_{\pm}$ depend only on the fields $A_\pm,B_\pm$ of the respective sign index, leading to vanishing rows and columns in the individual Hessians $\big(\Gamma^{\rm quad}_+\big)^{(2)}$ and $\big(\Gamma^{\rm quad}_-\big)^{(2)}$, that correspond to the fields of the other sign index. Using the decomposition \eqref{selfdual_decomp} we did not restrict the field space of fluctuations to one chirality. This restriction can be carried out at this stage by simply discarding the (vanishing) rows and columns of the Hessian that correspond to the other chirality. At the same time, the trace 'STr' is restricted to the subspace orthogonal to that of the deleted rows and columns. Thus we have a simple method at hand that reduces the $40 \times 40$ matrix operator of QECG to the $28 \times 28$ Hessian of (anti-)selfdual gravity, reflecting the reduced number of independent field components in the spin connection. The reduction of the \Ofour-ghosts proceeds in complete analogy. 

\subsection{Derivation of the $\beta$-functions}\label{Beta_Functions}
We are now in the position to derive the $\beta$-functions of Newton's constant and the cosmological constant in (anti-)selfdual gravity. We thus start with an action of the form 
\begin{equation}
\Gamma^\pm_k=-\frac{1}{8\pi G_k}\int {\rm d}^4 x e \Big[e_a{}^\mu e_b{}^\nu F(\omega^{(\pm)})^{ab}{}_{\mu\nu}- \Lambda_k\Big]+\Gamma^\pm_{\rm gf} +S^\pm_{\rm gh},
\end{equation}
which corresponds to the Holst truncation with $\gamma=\mp 1$ and the gauge fixing and ghost terms modified as discussed in the last subsection. Then, the left hand side of the flow equation reads
\begin{equation}
\begin{aligned}
\partial_t \Gamma^\pm_k[\bar e,\bar \omega^\pm]=&-\frac{k^2}{8\pi g_k}\bigg(2-\frac{\partial_t g_k}{g_k}\bigg) & &\hspace{-0.3cm}\cdot\!\int{\rm d}^d x\, \bar{e}\, \bar{e}_a{}^\mu\bar{e}_b{}^\nu \bar{F}(\omega^{(\pm)})^{ab}{}_{\mu\nu} \\
&+\frac{k^2}{8 \pi g_k}\bigg(\!2\!-\!\frac{\partial_t g_k}{g_k}\!+\!2\!+\!\frac{\partial_t \lambda_k}{\lambda_k}\bigg)\lambda_k k^2 & &\hspace{-0.3cm}\cdot\! \int {\rm d}^d x \,\bar{e}\:.
\end{aligned}
\end{equation}
Inserting the constant background fields $\bar{e}$ and $\bar{\omega}^{(\pm)}$ we will identify the prefactor of the field strength term on the right hand side, denoted rhsF{}$^\pm$, by the combination of $\big(\bar{\omega}^{(\pm)}\big)^2$-contractions:
\begin{equation}
  \bar{e}\, \bar{e}_a^{\ \mu}\bar{e}_b^{\ \nu} \bar{F}^{ab}_{\ \ \mu\nu}=\bar{e}\,\Big[\big(\bar{\omega}^{(\pm)}\big)_{abc}\big(\bar{\omega}^{(\pm)}\big)^{acb}-\big(\bar{\omega}^{(\pm)}\big)^{a}_{\ ca}\big(\bar{\omega}^{(\pm)}\big)^{bc}_{\ \ b}\Big]\:.
\end{equation}
As for the non-selfdual Holst truncation \cite{long}, this cannot be done unambiguously: For an (anti-)selfdual background spin connection, {\it any} contraction quadratic in $\bar{\omega}^{(\pm)}$ can be shown to be equal to the above two, but {\it with a different relative weight}\footnote{The proof parallels Appendix A.2 in \cite{long}.} (of the 5 independent torsion squared invariants of the general case, on an (anti-)selfdual background only two remain linearly independent). Hence, we need to specify exactly one additional basis vector on theory space besides the curvature term in order to identify its prefactor unambiguously.

Following this reasoning we evaluate the RHS of the flow equation and finally cast it into the form
\begin{equation}
\begin{aligned}
\partial_t \Gamma^\pm_k=& {\rm rhsF}^\pm \cdot k^2 \int\! {\rm d}^d x\, \bar{e}\, \Big(\big(\bar{\omega}^{(\pm)}\big)_{abc}\big(\bar{\omega}^{(\pm)}\big)^{acb}-\big(\bar{\omega}^{(\pm)}\big)^{a}_{\ ca}\big(\bar{\omega}^{(\pm)}\big)^{bc}_{\ \ b}\Big)\\ &+{\rm rhs\Lambda^\pm}\cdot k^4 \int\! {\rm d}^d x\,\bar{e} + {\rm rhsI}_\varphi^{\pm}\cdot k^2 \int \!{\rm d}^d x \, \bar{e} \,\bar{I}^{(\pm)}_\varphi\:.
\end{aligned}
\end{equation}
Here, $\bar{I}^{(\pm)}_\varphi$ is the additional vector that completes the basis in the projected part of theory space. Concretely we employ the following one parameter family of $(\bar{\omega}^{(\pm)})^2$-contractions:
\begin{equation}
\bar{I}^{(\pm)}_\varphi= \sin(\varphi)\big(\bar{\omega}^{(\pm)}\big)^{abc}\big(\bar{\omega}^{(\pm)}\big)_{acb}+\cos(\varphi)\big(\bar{\omega}^{(\pm)}\big)^{ab}{}_a\big(\bar{\omega}^{(\pm)}\big)^{c}{}_{bc}
\end{equation}
While we are not actually interested in the correponding prefactor, ${\rm rhsI}_\varphi^{\pm}$, the prefactor ${\rm rhsF}^\pm $ which enters the $\beta$-functions of $G_k$ and $\Lambda_k$ will depend on the value of $\varphi$ in general.\footnote{See Section 5 of ref. \cite{long} for a detailed discussion of this issue.} 

For a general choice of gauge parameters we find that rhsF${}^\pm(\lambda)$, rhsI${}^\pm_i(\lambda)$ only depend on the cosmological constant $\lambda$, and that these functions in $\lambda$ are given as the ratio of two polynomials of degree 10, with a common denominator. In the $(\alpha_{\rm D},\alpha_{\rm L}',\beta_{\rm D})=(0,0,0)$-gauge these polynomials simplify, such that the remainder is a ratio of polynomials of degree 4. Unfortunately, the simplification is not as impressive as in the case of the full Holst truncation, where a reduction to degree 1 was obtained. Nonetheless this gauge leads to the most extensive simplification possible and we will thus stick to the $(0,0,0)$-gauge in the following. 

\begin{figure}[ht]
\centering
{\small
\begin{psfrags}
 \input{LimitingValue-psfrag.tex}
 \includegraphics[width=0.65\linewidth]{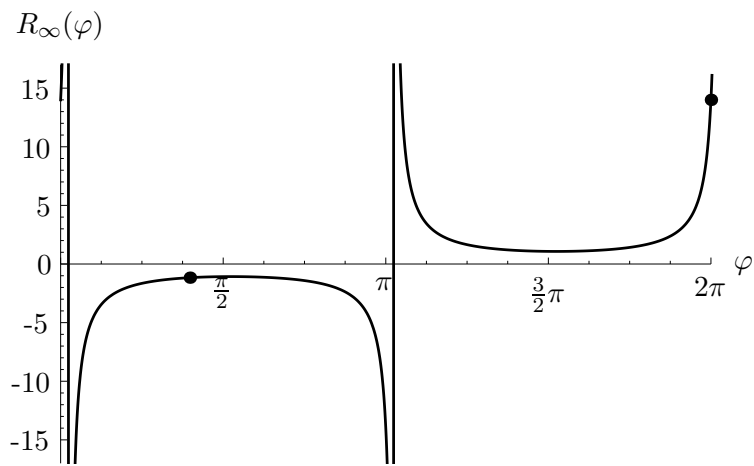}
\end{psfrags}
}
\caption{Asymptotic ratio of the coordinate functions $R_{\infty}=\lim_{\lambda\rightarrow \infty}{\rm rhsF}^\pm/{\rm rhs}\bar{I}^\pm_\varphi$ as a function of the basis parameter $\varphi$.}
\label{LimitingValue}
\end{figure}

In order to judge the reliability of the different choices of bases we could monitor the $\varphi$-dependence of the ratio rhsF${}^\pm(\lambda)$/rhsI${}^\pm_\varphi(\lambda)$, as in the QECG case. As an example, Fig. \ref{LimitingValue} shows the limiting value $R_{\infty}(\varphi)=\lim_{\lambda\rightarrow \infty}$ rhsF${}^\pm(\lambda)$/rhsI${}^\pm_\varphi(\lambda)$ that had been considered in the non-selfdual case already. 

As the main result of this subsection let us now write down the $\beta$-functions obtained for (anti-)selfdual gravity. For the dimensionless couplings $g_k\equiv k^2 G_k$ and $\lambda_k\equiv \Lambda_k/k^2$ they read

\begin{equation}
\begin{aligned}\label{betas_chiral}
\beta_g(\lambda,g)&= +2g+ 8 \pi g^2 \:{\rm rhsF}^\pm(\lambda) \\
\beta_\lambda(\lambda,g)&= -2 \lambda + 8 \pi g \lambda\: {\rm rhsF}^\pm(\lambda)+8 \pi g \:{\rm rhs}\Lambda^\pm(\lambda,g)\:.
\end{aligned}
\end{equation}
In the $(0,0,0)$-gauge, the coefficient functions take on the form
\begin{align}
{\rm rhs}\Lambda^{\pm}(\lambda,g)&=-\frac{1}{32 \pi^2}\bigg(\ln \bigg[\frac{(\lambda-1)^{12}\lambda^6}{g^{50} m^{50}}\bigg]-\ln {\cal N^\pm}\bigg),\ {\rm with}\  \ln {\cal N^\pm}\approx 151.5\\
{\rm rhsF}^\pm(\lambda)&=-\frac{\left(-156 \lambda^4+223 \lambda^3+132 \lambda^2-136 \lambda +12\right) \sin (\varphi )}{512 \pi^2 (\lambda -1)^2 \lambda^2 (\sin (\varphi )+\cos (\varphi ))}\nonumber\\&\quad-\frac{\left(12
   \lambda^4-277 \lambda^3+244 \lambda^2-40 \lambda -4\right) \cos (\varphi )}{512 \pi^2 (\lambda -1)^2 \lambda^2 (\sin (\varphi )+\cos (\varphi ))}\label{rhsFpm}
\end{align}
Note that ${\rm rhs}\Lambda^{\pm}(\lambda,g)$ depends on both $\lambda$ and $g$, but it is independent of $\varphi$. Furthermore, it contains the normalization parameter $\mu\equiv \bar\mu/k^2$ whose role has been discussed in \cite{long} already. From now on we set it to its natural value $\mu=1$. Note also that the $\beta$-functions for the selfdual and the anti-selfdual case are exactly the same.

\subsection{Analysis of the RG flow}\label{RG_Flow}

In this subsection we are going to analyze the RG flow of selfdual gravity resulting from the system of differential equations $\partial_t g =\beta_g, \partial_t\lambda=\beta_\lambda$ with the $\beta$-functions \eqref{betas_chiral}, whose explicit form depends on the basis chosen, cf. eq. \eqref{rhsFpm}. 

A first look reveals a divergence of both $\beta$-functions on the line $\lambda=0$, which comes in addition to the divergence at $\lambda=1$, that is known already from the QECG case \cite{long}. For a generic choice of gauge parameters there would be even more divergences for fixed $\lambda$, all of which move to ``$\lambda=\infty$'' when approaching the $(\aD,\alpha_L,\bD)=(0,0,0)$ limit. The ``new'' divergence at $\lambda=0$ corresponds to a gauge dependent zero of the denominator that approaches zero in this limit. 

This divergence has an interesting effect: As the pole in $\beta_g$ is of one degree higher than the one in $\beta_\lambda$, the RG trajectories do not reach this line. Thus all trajectories in the $(g>0, \lambda>0)$-quadrant are confined to this quadrant. We will see that in the IR they either run to small values of $g$ and large values of $\lambda$ (which we know as type IIIa trajectories from metric gravity and QECG) or to small $\lambda$ and large values of $g$, which amounts to a completely new IR behavior seen for the first time in gravity. 

{\bf \noindent Fixed point structure.}
Since the origin of the $g$-$\lambda$-theory space now lies on a singular line, a Gaussian fixed point cannot be properly defined there.

However, we do find non-Gaussian fixed points (NGFPs) at fixed point values $(g^\ast, \lambda^\ast)$ by solving the condition $\beta_g(\lambda,g)=0$ for $g^\ast(\lambda)=-[4\pi\, {\rm rhsF}^\pm(\lambda)]^{-1}$, substituting this solution into the second condition $\beta_\lambda(\lambda, g^\ast(\lambda))$ $=0$ and searching for its zeros. This final step can only be carried out numerically, due to the logarithmic terms in rhs$\Lambda^\pm(\lambda)$. 

Doing this for the continuous set of bases labeled by $\varphi$, we generically found two fixed points in the range $\lambda<1$, one at small positive $\lambda$, which we will denote by ${\bf NGFP_\pm^1}$, and the second at large negative $\lambda$ (${\bf NGFP_\pm^2}$). This picture resembles very much the situation in any $\lambda$-$g$-plane of fixed $\gamma\neq \pm 1$ of the full Holst truncation. However, we find that the existence of the fixed points depends on the value of $\varphi$, \ie on the basis chosen. We will discuss this issue in more detail below, after having first analyzed the properties of the fixed points. 

Besides these two most stable fixed point solutions we found additional solutions, that were considered unphysical, as they occur very close to singularities of the function $\beta_\lambda(\lambda,g^\ast(\lambda))$ and and the influence on the RG flow of the fixed points they give rise to is very localized.

\begin{figure}[tb]
\centering
{\small
\subfigure[]{\label{NGFPpm1_pos}
\begin{psfrags}
 \input{NGFPpm1_pos-psfrag.tex}
 \includegraphics[width=0.45\linewidth]{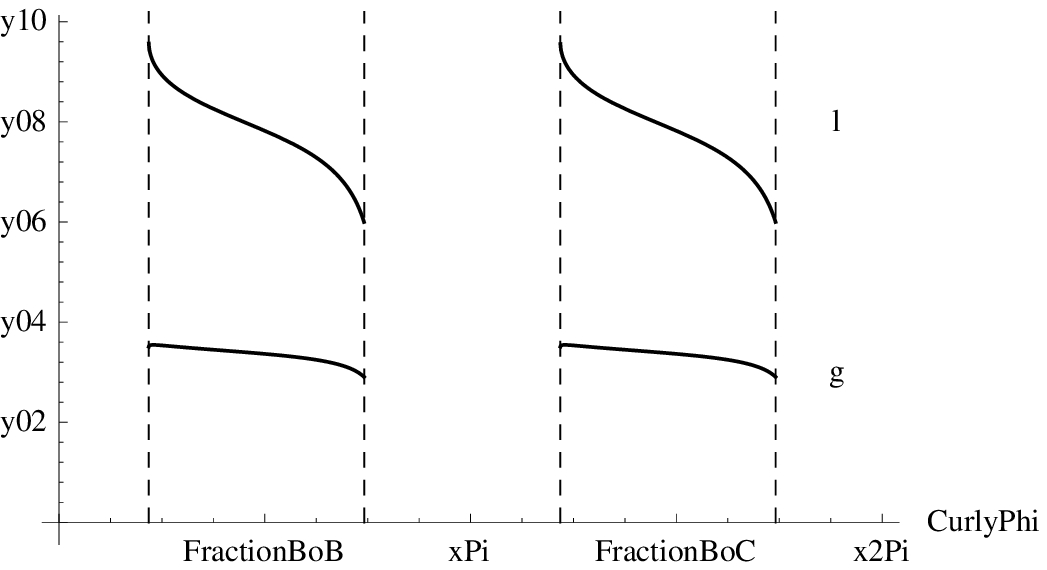}
\end{psfrags}}\quad
\subfigure[]{\label{NGFPpm1_CE}
\begin{psfrags}
 \input{NGFPpm1_CE-psfrag.tex}
 \includegraphics[width=0.45\linewidth]{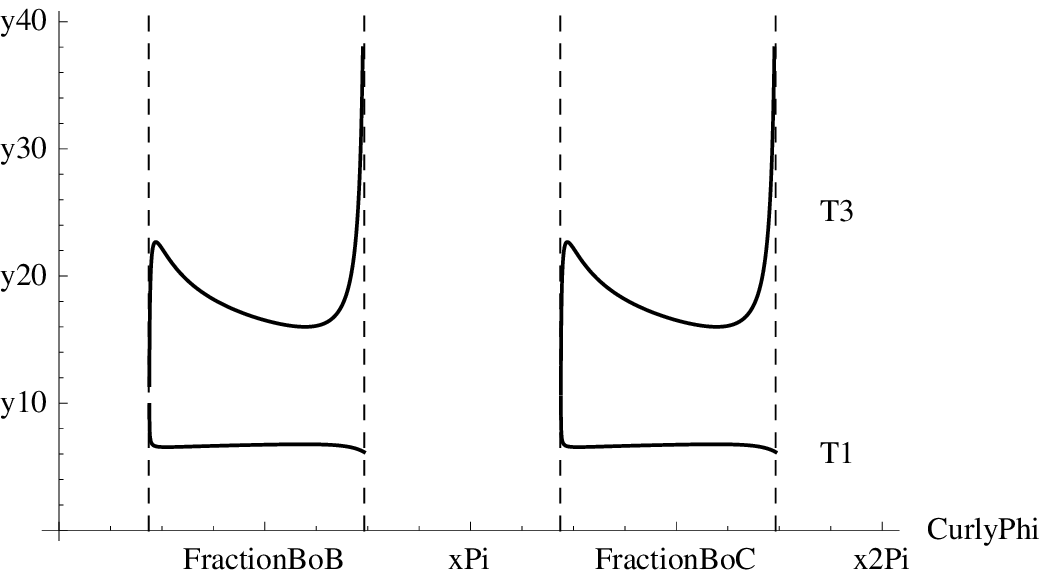}
\end{psfrags}}}
\caption{Coordinates and critical exponents of ${\bf NGFP_\pm^1}$ as a function of the parameter $\varphi$.}
\end{figure}

\noindent {\bf (A) The fixed point} ${\bf NGFP_\pm^1.}$ In Fig. \ref{NGFPpm1_pos} we have plotted the coordinates of the first NGFP as a function of the basis parameter $\varphi$. Since rhsI${}^\pm_i$ switches its sign, while rhsF${}^\pm$ stays constant under $\varphi\mapsto\varphi+\pi$, this and the following figures are $\pi$-periodic in $\varphi$. We observe that the fixed point is only present in the interval $\pi/4\lesssim\varphi\lesssim 3/4 \pi$ (and its $\pi$-periodic counterpart). We see that both $\lambda^\ast$ and $g^\ast$ decrease with increasing $\varphi$, but both coordinate values change by much less than one order of magnitude. In particular $g^\ast$ turns out remarkably stable, having a plateau value of about $g^\ast\approx 0.35$. 

Fig. \ref{NGFPpm1_CE} shows the $\varphi$-dependence of the critical exponents; following the conventions of \cite{long}, we denote them $\theta_1$ and $\theta_3$. At the lower boundary of the interval in which the fixed point exists, there seems to be a bifurcation point, where the critical exponents become real. Quickly thereafter $\theta_1$ approaches a quite stable plateau with a value of about $6.5$, while $\theta_3$ fluctuates around $20$, before it diverges at the upper boundary of the interval. Most importantly, both critical exponents are positive, such that the fixed point is UV attractive. 

Qualitatively, but also quantitatively this fixed point resembles much the fixed point ${\bf NGFP_{\boldsymbol{\infty}}^1}$ of the full Holst truncation. Either of them has fixed point coordinates that are smaller than unity and relatively stable, and one of their critical exponents takes on a fairly large value.
 
\begin{figure}[tb]
\centering
{\small
\subfigure[]{\label{NGFPpm2_pos}
\begin{psfrags}
 \input{NGFPpm2_pos-psfrag.tex}
 \includegraphics[width=0.45\linewidth]{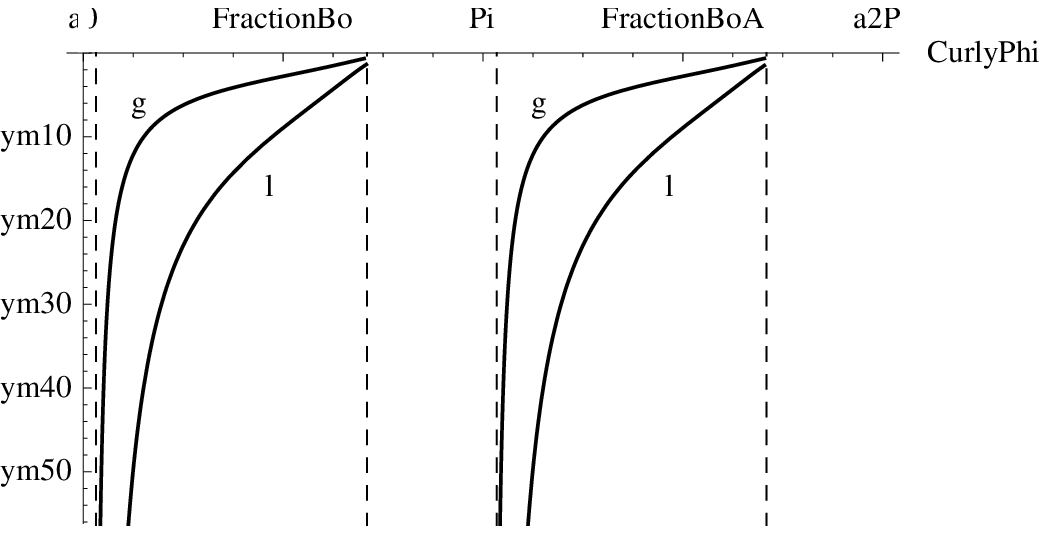}
\end{psfrags}}\quad
\subfigure[]{\label{NGFPpm2_CE}
\begin{psfrags}
 \input{NGFPpm2_CE-psfrag.tex}
 \includegraphics[width=0.45\linewidth]{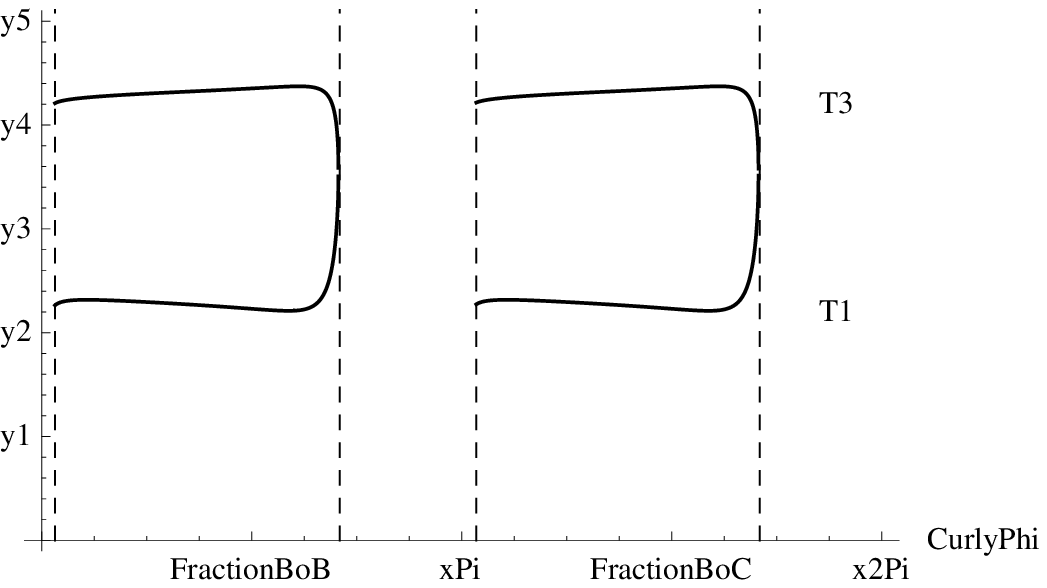}
\end{psfrags}}}
\caption{Coordinates and critical exponents of ${\bf NGFP_\pm^2}$ as a function of the parameter $\varphi$.}
\end{figure}

\noindent {\bf(B) The fixed point} ${\bf NGFP_\pm^2.}$ Let us turn over to the second fixed point. Its coordinates as a function of $\varphi$ are depicted in Fig. \ref{NGFPpm2_pos}. We observe that it exists for a slightly larger interval in $\varphi$: The lower boundary is shifted to $\varphi\approx 0.1$, while the upper boundary occurs at virtually the same value $\varphi\approx 3/4 \pi$ as in the case of ${\bf NGFP_\pm^1}$. We find that the fixed point position heavily depends on the value of $\varphi$: It starts at infinite negative values at the lower boundary and moves close to the origin at the upper boundary. In between it always stays within the $(\lambda<0,g<0)$ quadrant.

The corresponding critical exponents are depicted in Fig. \ref{NGFPpm2_CE}. Their almost perfect independence on $\varphi$, taking into account the huge variation of the fixed point position is most striking: Both critical exponents are approximately constant with $\theta_1\approx2.3$ and $\theta_3\approx 4.3$. In particular, both exponents are real and positive, giving rise to a UV attractivity of the FP in both directions. 

Also the properties of ${\bf NGFP_\pm^2}$ suggest that it is a descendant of a fixed point in the non-selfdual theory: They all are comparable to those of ${\bf NGFP_{\boldsymbol{\infty}}^2}$ found in the full Holst truncation \cite{long}. 

\noindent{\bf (C) Discussion.} Let us finally comment in more detail on the interval of existence of both fixed points. Naively one could think that universal properties of the flow, like the existence of fixed points, should also be independent of the basis chosen. This is not true, as the projection from the full theory space onto the truncated subspace clearly can be chosen in a particularly disadvantageous way, such that the physical content of the theory is projected out. While it is impossible to identify the best projection of the flow without knowing its exact untruncated form, in the case at hand we know two of these disadvantageous choices for $\varphi$: 

\noindent{\bf (i)} Those $\varphi$ at which the poles in Fig. \ref{LimitingValue} occur, correspond to a basis, where the second invariant points exactly into the direction of the expanded RHS of the flow equation. Hence, rhsF${}^\pm$ vanishes in this case and the information we are interested in is projected out.

\noindent{\bf(ii)} At $\varphi=3/4 \pi$ both basis vectors point in the same direction, \ie are linearly dependent. Thus, in this limit, both coordinate functions rhsF${}^\pm$ and rhs$\bar{I}^\pm_\varphi$ diverge and, although their ratio stays finite, the extracted RG flow becomes questionable.

It is certainly no mere coincidence that the boundaries of the interval of existence of ${\bf NGFP_{\boldsymbol{\infty}}^2}$ (and also the upper boundary for ${\bf NGFP_{\boldsymbol{\infty}}^1}$) lie very close to these extreme cases. From this point of view one should consider a basis in the middle of this interval as most reliable. For the phase portrait we shall therefore use $\varphi=\arctan{3}\approx 0.4 \pi$ as a representative value.

\begin{figure}[ht]
\centering
{\small
\subfigure[]{
\begin{psfrags}
 \input{Chiral_PhasePortrait1-psfrag.tex}
 \includegraphics[width=0.45\linewidth]{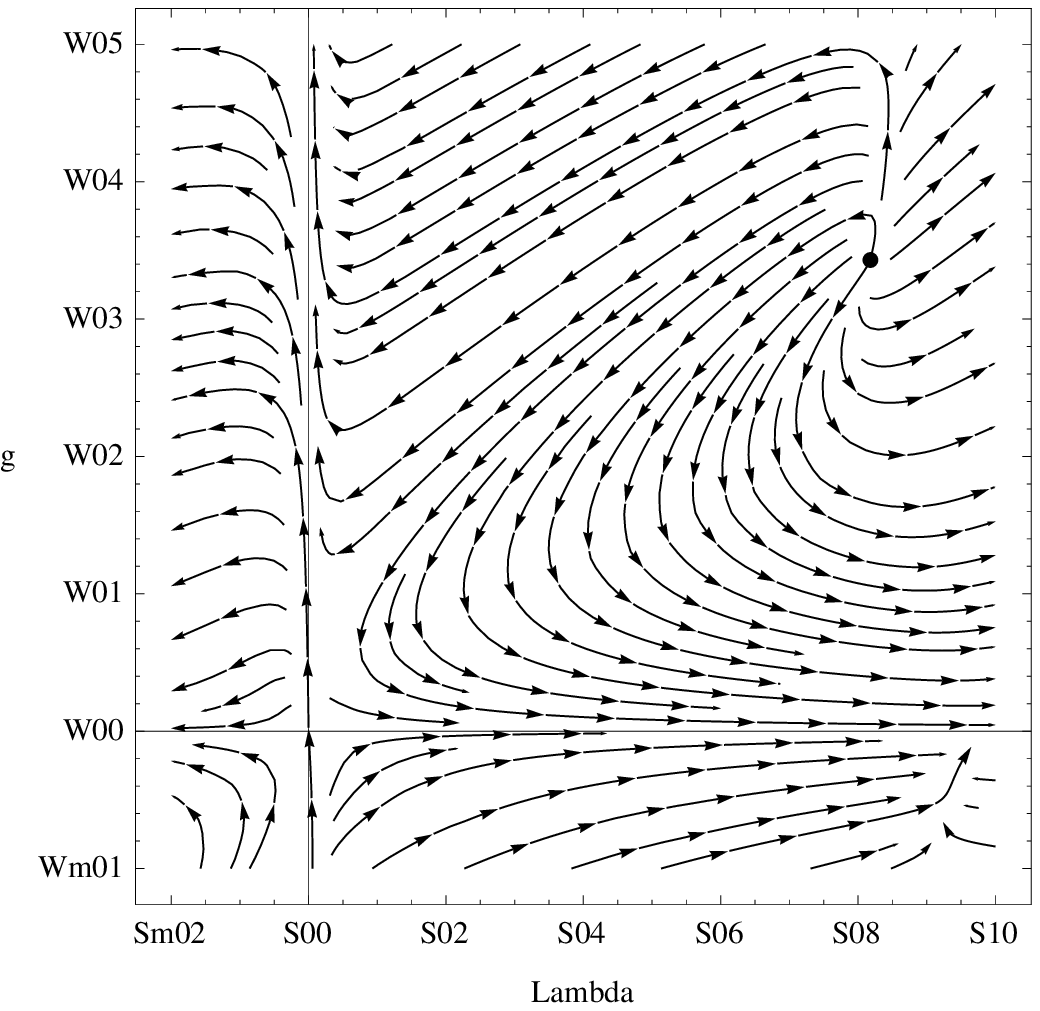}
\end{psfrags}}\quad
\subfigure[]{
\begin{psfrags}
 \input{Chiral_PhasePortrait2-psfrag.tex}
 \includegraphics[width=0.45\linewidth]{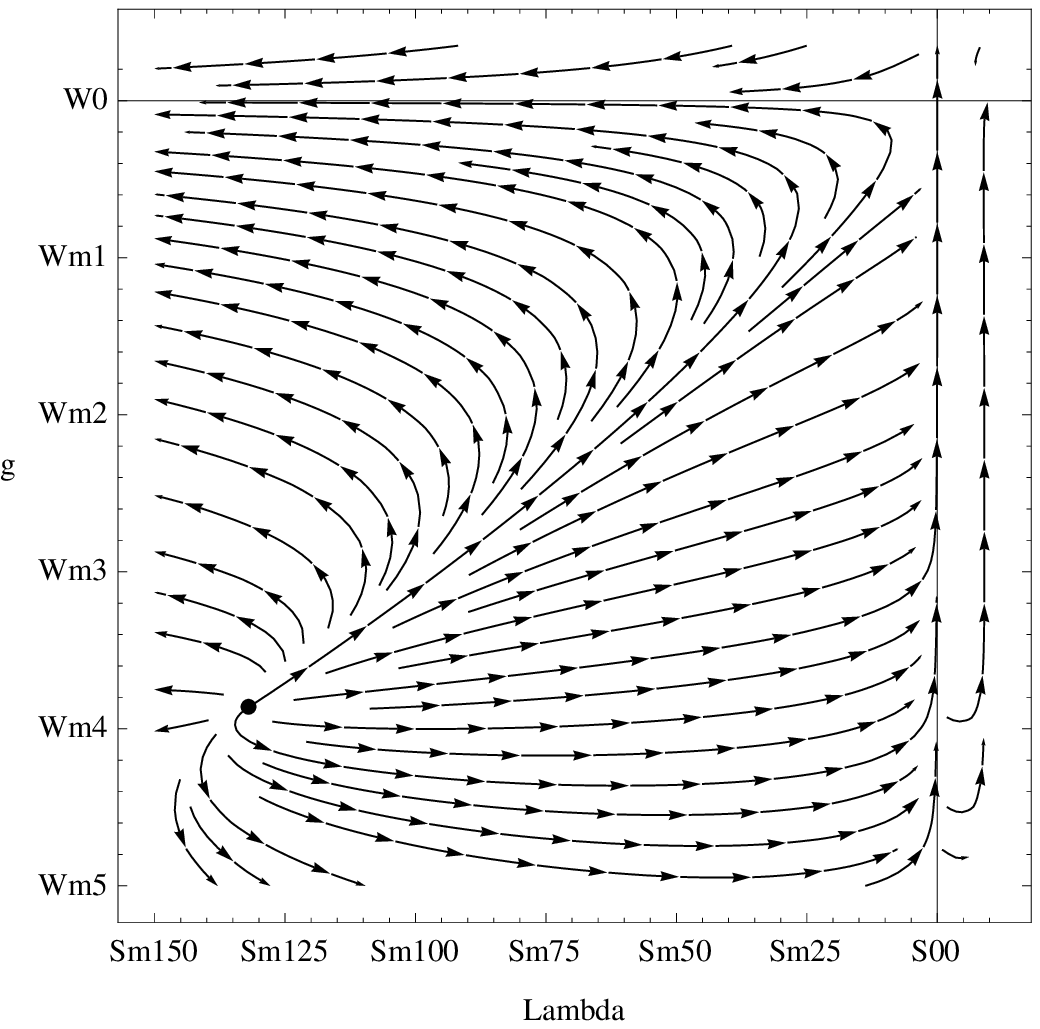}
\end{psfrags}}}
\caption{Phase portrait of selfdual gravity.}
\label{Chiral_PhasePortrait}
\end{figure}

{\bf \noindent The phase portrait.}
In Fig. \ref{Chiral_PhasePortrait} we plot the phase portrait of the RG flow of selfdual gravity. 
In subfigure (a) the vicinity of ${\bf NGFP_\pm^1}$ and the flow towards the origin is depicted. We observe that the trajectories shortly before arriving at the origin are bent to one side or the other, such that they either run towards large values of $\lambda$ and small $g$ (as known from metric gravity) or to large $g$ and small $\lambda$ in the IR. This new behavior is clearly due to the existence of the additional divergence at $\lambda=0$ compared to both metric gravity and the QECG case.

Subfigure (b) focusses on ${\bf NGFP_\pm^2}$ and the $(\lambda<0,g<0)$ quadrant. It shows no particular differences compared to the $(\lambda,g)$-truncations for fixed $\gamma\neq\pm1$ of the Holst action, except for the divergence at $\lambda=0$.

This additional divergence, however, should probably not be taken too seriously. In fact, we were able to trace back its origin to the modified gauge condition ${\cal G}^\pm_{ab}$ in eq. \eqref{ChiralGF}. Picking the ``+'' chirality for the sake of the argument, it leads to a gauge fixing action $\Gamma^{\rm gf}_{k}$ containing $ {\cal G}^+_{ab}{\cal G}^{+\,ab}=(P^+{\cal G})_{ab}(P^+{\cal G})^{ab}$. However it should also be admissible to use the {\it complete} gauge condition ${\cal G}_{ab}$ in the gauge fixing action $S^\pm_{\rm gf}$. It would decompose according to ${\cal G}_{ab}{\cal G}^{ab}=(P^+{\cal G})_{ab}(P^+{\cal G})^{ab}+(P^-{\cal G})_{ab}(P^-{\cal G})^{ab}$, where the second term on the RHS is simply invariant under selfdual, \ie ${\sf SO}^+(3)_{\rm loc}$ transformations, while the first still gauge fixes them. Using this second gauge condition, {\it the divergence at $\lambda=0$ is no longer present in the $\beta$-functions}. However, it has the disadvantage that it is not possible to take the limit of the preferred $(0,0,0)$-gauge in this case. For that reason we opted for the chiral gauge condition, which irrespective of the practical considerations seems the most natural choice. Nonetheless this observation puts the physical meaning of the divergence arising at $\lambda=0$ into question.

Taking together all our findings on the RG flow of selfdual gravity, we conclude that setting $\gamma=\mp1$ results in a self-consistent ``sub-truncation'' within the general Holst action ansatz. Most strikingly the resulting phase portrait and the properties of the two NGFPs we found correspond, qualitatively and quantitatively, very well to the other self-consistent sub-truncation, namely the $1/\gamma=0$-plane, in which the Immirzi parameter is not renormalized, too.

\section{Summary}
In this letter we employed a description of Euclidean gravity in 4 dimensions which involves a selfdual or anti-selfdual spin-connection, $\omega^{(\pm) ab}_{\ \ \ \ \mu}$ alongside with the tetrad field, $e^a_{\ \mu}$. It gives rise to a theory space of action functionals, $\{A[e, \omega^{(\pm)}]\}$, which, when suitably generalized by background and ghost fields, can support a non-perturbative coarse graining flow. Trying to get a first understanding of the RG flow on this new space we took advantage of a ``special purpose'' functional RG equation that has been constructed recently for the closely related theory space pertaining to (non-selfdual) Einstein-Cartan gravity \cite{long}. In ref. \cite{long} we had computed the RG flow implied by a scale dependent Holst action essentially. The natural coordinates on this 3-dimensional truncated theory space are triples $(g, \lambda, \gamma)$ whereby $\gamma=\pm 1$ corresponds to two singular planes on which the Einstein-Cartan FRGE breaks down since the spin connections lose half of their independent field components and become (anti-)selfdual there. As a consequence, the present investigation using selfdual connections is by no means a ``special case'' of the general Einstein-Cartan setting in which the Immirzi parameter was allowed to run. In principle selfdual and Einstein-Cartan gravity may have entirely different RG properties, being based on disconnected theory spaces. Nevertheless, this is not what actually seems to happen: Here we found little to no qualitative difference of the selfdual flow when compared to the Einstein-Cartan flow of $g$ and $\lambda$ in planes of constant $\gamma\neq \pm 1$. Especially the $(g>0, \lambda>0)$-quadrant shows a striking similarity to the Einstein-Hilbert truncation for metric gravity, although the critical exponents of the non-Gaussian fixed point it contains are real in the present case and their absolute value is larger.

Thus we may conclude that the theory space with (anti-)selfdual connections is likely to be asymptotically safe, too. In fact, the evidence for its non-perturbative renormalizability is about as strong as for Einstein-Cartan gravity without the selfduality condition.


\end{document}

%% file: LimitingValue-psfrag.tex
\def\PFGstripminus-#1{#1}%
\def\PFGshift(#1,#2)#3{\raisebox{#2}[\height][\depth]{\hbox{%
  \ifdim#1<0pt\kern#1 #3\kern\PFGstripminus#1\else\kern#1 #3\kern-#1\fi}}}%
\providecommand{\PFGstyle}{}%
%
\psfrag{CurlyPhi}[cl][cl]{\PFGstyle $\varphi $}%
\psfrag{FractionBoA}[Bc][Bc][1.][0.]{\PFGstyle $\frac{3}{2}\pi$}%
\psfrag{FractionBo}[Bc][Bc][1.][0.]{\PFGstyle $\frac{\pi}{2}$}%
\psfrag{FractionBoB}[tc][tc]{\PFGstyle $\frac{\pi}{2}$}%
\psfrag{FractionBoC}[tc][tc]{\PFGstyle $\frac{3}{2}\pi$}%
\psfrag{FractionBoD}[Bc][Bc]{\PFGstyle $\frac{\pi}{2}$}%
\psfrag{FractionBoE}[Bc][Bc]{\PFGstyle $\frac{3}{2}\pi$}%
\psfrag{RinfA}[Bc][Bc]{\PFGstyle $R_\infty(\varphi)$}%
\psfrag{Rinf}[bc][bc]{\PFGstyle $R_\infty(\varphi)$}%
\psfrag{x0}[tc][tc]{\PFGstyle $0$}%
\psfrag{x2Pi}[tc][tc]{\PFGstyle $\text{2$\pi $}$}%
\psfrag{xPi}[tc][tc]{\PFGstyle $\pi $}%
\psfrag{y0}[cr][cr]{\PFGstyle $\text{ 0}$}%
\psfrag{y10}[cr][cr]{\PFGstyle $10$}%
\psfrag{y15}[cr][cr]{\PFGstyle $15$}%
\psfrag{y5}[cr][cr]{\PFGstyle $\text{ 5}$}%
\psfrag{ym10}[cr][cr]{\PFGstyle $\text{-10}$}%
\psfrag{ym15}[cr][cr]{\PFGstyle $\text{-15}$}%
\psfrag{ym5}[cr][cr]{\PFGstyle $\text{ -5}$}%

%% file: NGFPpm1_pos-psfrag.tex
\def\PFGstripminus-#1{#1}%
\def\PFGshift(#1,#2)#3{\raisebox{#2}[\height][\depth]{\hbox{%
  \ifdim#1<0pt\kern#1 #3\kern\PFGstripminus#1\else\kern#1 #3\kern-#1\fi}}}%
\providecommand{\PFGstyle}{}%
%
\psfrag{CurlyPhi}[cl][cl]{\PFGstyle $\varphi $}%
\psfrag{FractionBoA}[Bc][Bc][1.][0.]{\PFGstyle $\frac{3}{2}\pi$}%
\psfrag{FractionBo}[Bc][Bc][1.][0.]{\PFGstyle $\frac{\pi}{2}$}%
\psfrag{FractionBoB}[tc][tc]{\PFGstyle $\frac{\pi}{2}$}%
\psfrag{FractionBoC}[tc][tc]{\PFGstyle $\frac{3}{2}\pi$}%
\psfrag{FractionBoD}[Bc][Bc]{\PFGstyle $\frac{\pi}{2}$}%
\psfrag{FractionBoE}[Bc][Bc]{\PFGstyle $\frac{3}{2}\pi$}%
\psfrag{g}[cc][cc]{\PFGstyle $g^\ast$}%
\psfrag{l}[cc][cc]{\PFGstyle $\lambda^\ast$}%
\psfrag{x0}[tc][tc]{\PFGstyle $0$}%
\psfrag{x2Pi}[tc][tc]{\PFGstyle $\text{2$\pi $}$}%
\psfrag{xPi}[tc][tc]{\PFGstyle $\pi $}%
\psfrag{y00}[cr][cr]{\PFGstyle $0.0$}%
\psfrag{y02}[cr][cr]{\PFGstyle $0.2$}%
\psfrag{y04}[cr][cr]{\PFGstyle $0.4$}%
\psfrag{y06}[cr][cr]{\PFGstyle $0.6$}%
\psfrag{y08}[cr][cr]{\PFGstyle $0.8$}%
\psfrag{y10}[cr][cr]{\PFGstyle $1.0$}%

%% file: NGFPpm1_CE-psfrag.tex
\def\PFGstripminus-#1{#1}%
\def\PFGshift(#1,#2)#3{\raisebox{#2}[\height][\depth]{\hbox{%
  \ifdim#1<0pt\kern#1 #3\kern\PFGstripminus#1\else\kern#1 #3\kern-#1\fi}}}%
\providecommand{\PFGstyle}{}%
%
\psfrag{CurlyPhi}[cl][cl]{\PFGstyle $\varphi $}%
\psfrag{FractionBoA}[Bc][Bc][1.][0.]{\PFGstyle $\frac{3}{2}\pi$}%
\psfrag{FractionBo}[Bc][Bc][1.][0.]{\PFGstyle $\frac{\pi}{2}$}%
\psfrag{FractionBoB}[tc][tc]{\PFGstyle $\frac{\pi}{2}$}%
\psfrag{FractionBoC}[tc][tc]{\PFGstyle $\frac{3}{2}\pi$}%
\psfrag{FractionBoD}[Bc][Bc]{\PFGstyle $\frac{\pi}{2}$}%
\psfrag{FractionBoE}[Bc][Bc]{\PFGstyle $\frac{3}{2}\pi$}%
\psfrag{T1}[cc][cc]{\PFGstyle $\theta_1$}%
\psfrag{T3}[cc][cc]{\PFGstyle $\theta_3$}%
\psfrag{x0}[tc][tc]{\PFGstyle $0$}%
\psfrag{x2Pi}[tc][tc]{\PFGstyle $\text{2$\pi $}$}%
\psfrag{xPi}[tc][tc]{\PFGstyle $\pi $}%
\psfrag{y0}[cr][cr]{\PFGstyle $\text{ 0}$}%
\psfrag{y10}[cr][cr]{\PFGstyle $10$}%
\psfrag{y20}[cr][cr]{\PFGstyle $20$}%
\psfrag{y30}[cr][cr]{\PFGstyle $30$}%
\psfrag{y40}[cr][cr]{\PFGstyle $40$}%

%% file: NGFPpm2_pos-psfrag.tex
\def\PFGstripminus-#1{#1}%
\def\PFGshift(#1,#2)#3{\raisebox{#2}[\height][\depth]{\hbox{%
  \ifdim#1<0pt\kern#1 #3\kern\PFGstripminus#1\else\kern#1 #3\kern-#1\fi}}}%
\providecommand{\PFGstyle}{}%
%
\psfrag{a0}[cc][cc]{\PFGstyle $0$}%
\psfrag{a2Pi}[cc][cc]{\PFGstyle $\text{2$\pi $}$}%
\psfrag{CurlyPhi}[cl][cl]{\PFGstyle $\varphi $}%
\psfrag{FractionBoA}[cc][cc][1.][0.]{\PFGstyle $\frac{3}{2}\pi$}%
\psfrag{FractionBoB}[Bc][Bc]{\PFGstyle $\frac{\pi}{2}$}%
\psfrag{FractionBoC}[Bc][Bc]{\PFGstyle $\frac{3}{2}\pi$}%
\psfrag{FractionBo}[cc][cc][1.][0.]{\PFGstyle $\frac{\pi}{2}$}%
\psfrag{g}[cc][cc]{\PFGstyle $g^\ast$}%
\psfrag{l}[cc][cc]{\PFGstyle $\lambda^\ast$}%
\psfrag{Pi}[cc][cc]{\PFGstyle $\pi $}%
\psfrag{x0A}[tc][tc]{\PFGstyle $0$}%
\psfrag{x0}[tc][tc][1.][0.]{\PFGstyle $0$}%
\psfrag{x11A}[tc][tc]{\PFGstyle $1$}%
\psfrag{x11}[tc][tc][1.][0.]{\PFGstyle $1$}%
\psfrag{x25A}[tc][tc]{\PFGstyle $0.25$}%
\psfrag{x25}[tc][tc][1.][0.]{\PFGstyle $0.25$}%
\psfrag{x5A}[tc][tc]{\PFGstyle $0.5$}%
\psfrag{x5}[tc][tc][1.][0.]{\PFGstyle $0.5$}%
\psfrag{x75A}[tc][tc]{\PFGstyle $0.75$}%
\psfrag{x75}[tc][tc][1.][0.]{\PFGstyle $0.75$}%
\psfrag{xm11A}[tc][tc]{\PFGstyle $-1$}%
\psfrag{xm11}[tc][tc][1.][0.]{\PFGstyle $-1$}%
\psfrag{xm25A}[tc][tc]{\PFGstyle $-0.25$}%
\psfrag{xm25}[tc][tc][1.][0.]{\PFGstyle $-0.25$}%
\psfrag{xm5A}[tc][tc]{\PFGstyle $-0.5$}%
\psfrag{xm5}[tc][tc][1.][0.]{\PFGstyle $-0.5$}%
\psfrag{xm75A}[tc][tc]{\PFGstyle $-0.75$}%
\psfrag{xm75}[tc][tc][1.][0.]{\PFGstyle $-0.75$}%
\psfrag{y0}[cr][cr]{\PFGstyle $\text{ 0}$}%
\psfrag{y11A}[cr][cr]{\PFGstyle $1$}%
\psfrag{y11}[cr][cr][1.][0.]{\PFGstyle $1$}%
\psfrag{y21A}[cr][cr]{\PFGstyle $2$}%
\psfrag{y21}[cr][cr][1.][0.]{\PFGstyle $2$}%
\psfrag{y31A}[cr][cr]{\PFGstyle $3$}%
\psfrag{y31}[cr][cr][1.][0.]{\PFGstyle $3$}%
\psfrag{y41A}[cr][cr]{\PFGstyle $4$}%
\psfrag{y41}[cr][cr][1.][0.]{\PFGstyle $4$}%
\psfrag{y51A}[cr][cr]{\PFGstyle $5$}%
\psfrag{y51}[cr][cr][1.][0.]{\PFGstyle $5$}%
\psfrag{y61A}[cr][cr]{\PFGstyle $6$}%
\psfrag{y61}[cr][cr][1.][0.]{\PFGstyle $6$}%
\psfrag{ym10}[cr][cr]{\PFGstyle $\text{-10}$}%
\psfrag{ym20}[cr][cr]{\PFGstyle $\text{-20}$}%
\psfrag{ym30}[cr][cr]{\PFGstyle $\text{-30}$}%
\psfrag{ym40}[cr][cr]{\PFGstyle $\text{-40}$}%
\psfrag{ym50}[cr][cr]{\PFGstyle $\text{-50}$}%
\psfrag{ym60}[cr][cr]{\PFGstyle $\text{-60}$}%

%% file: NGFPpm2_CE-psfrag.tex
\def\PFGstripminus-#1{#1}%
\def\PFGshift(#1,#2)#3{\raisebox{#2}[\height][\depth]{\hbox{%
  \ifdim#1<0pt\kern#1 #3\kern\PFGstripminus#1\else\kern#1 #3\kern-#1\fi}}}%
\providecommand{\PFGstyle}{}%
%
\psfrag{CurlyPhi}[cl][cl]{\PFGstyle $\varphi $}%
\psfrag{FractionBoA}[Bc][Bc][1.][0.]{\PFGstyle $\frac{3}{2}\pi$}%
\psfrag{FractionBo}[Bc][Bc][1.][0.]{\PFGstyle $\frac{\pi}{2}$}%
\psfrag{FractionBoB}[tc][tc]{\PFGstyle $\frac{\pi}{2}$}%
\psfrag{FractionBoC}[tc][tc]{\PFGstyle $\frac{3}{2}\pi$}%
\psfrag{FractionBoD}[Bc][Bc]{\PFGstyle $\frac{\pi}{2}$}%
\psfrag{FractionBoE}[Bc][Bc]{\PFGstyle $\frac{3}{2}\pi$}%
\psfrag{T1}[cc][cc]{\PFGstyle $\theta_1$}%
\psfrag{T3}[cc][cc]{\PFGstyle $\theta_3$}%
\psfrag{x0}[tc][tc]{\PFGstyle $0$}%
\psfrag{x2Pi}[tc][tc]{\PFGstyle $\text{2$\pi $}$}%
\psfrag{xPi}[tc][tc]{\PFGstyle $\pi $}%
\psfrag{y0}[cr][cr]{\PFGstyle $0$}%
\psfrag{y1}[cr][cr]{\PFGstyle $1$}%
\psfrag{y2}[cr][cr]{\PFGstyle $2$}%
\psfrag{y3}[cr][cr]{\PFGstyle $3$}%
\psfrag{y4}[cr][cr]{\PFGstyle $4$}%
\psfrag{y5}[cr][cr]{\PFGstyle $5$}%

%% file: Chiral_PhasePortrait1-psfrag.tex
\def\PFGstripminus-#1{#1}%
\def\PFGshift(#1,#2)#3{\raisebox{#2}[\height][\depth]{\hbox{%
  \ifdim#1<0pt\kern#1 #3\kern\PFGstripminus#1\else\kern#1 #3\kern-#1\fi}}}%
\providecommand{\PFGstyle}{}%
%
\psfrag{g}[cr][cr]{\PFGstyle $g$}%
\psfrag{Lambda}[tc][tc]{\PFGstyle $\lambda $}%
\psfrag{S00}[tc][tc]{\PFGstyle $0.0$}%
\psfrag{S02}[tc][tc]{\PFGstyle $0.2$}%
\psfrag{S04}[tc][tc]{\PFGstyle $0.4$}%
\psfrag{S06}[tc][tc]{\PFGstyle $0.6$}%
\psfrag{S08}[tc][tc]{\PFGstyle $0.8$}%
\psfrag{S10}[tc][tc]{\PFGstyle $1.0$}%
\psfrag{Sm02}[tc][tc]{\PFGstyle $\text{-0.2}$}%
\psfrag{W00}[cr][cr]{\PFGstyle $0.0$}%
\psfrag{W01}[cr][cr]{\PFGstyle $0.1$}%
\psfrag{W02}[cr][cr]{\PFGstyle $0.2$}%
\psfrag{W03}[cr][cr]{\PFGstyle $0.3$}%
\psfrag{W04}[cr][cr]{\PFGstyle $0.4$}%
\psfrag{W05}[cr][cr]{\PFGstyle $0.5$}%
\psfrag{Wm01}[cr][cr]{\PFGstyle $\text{-0.1}$}%
\psfrag{y00}[cr][cr]{\PFGstyle $0.0$}%
\psfrag{y01}[cr][cr]{\PFGstyle $0.1$}%
\psfrag{y02}[cr][cr]{\PFGstyle $0.2$}%
\psfrag{y03}[cr][cr]{\PFGstyle $0.3$}%
\psfrag{y04}[cr][cr]{\PFGstyle $0.4$}%
\psfrag{y05}[cr][cr]{\PFGstyle $0.5$}%
\psfrag{ym01}[cr][cr]{\PFGstyle $\text{-0.1}$}%

%% file: Chiral_PhasePortrait2-psfrag.tex
\def\PFGstripminus-#1{#1}%
\def\PFGshift(#1,#2)#3{\raisebox{#2}[\height][\depth]{\hbox{%
  \ifdim#1<0pt\kern#1 #3\kern\PFGstripminus#1\else\kern#1 #3\kern-#1\fi}}}%
\providecommand{\PFGstyle}{}%
%
\psfrag{g}[cr][cr]{\PFGstyle $g$}%
\psfrag{Lambda}[tc][tc]{\PFGstyle $\lambda $}%
\psfrag{S00}[tc][tc]{\PFGstyle $\text{0.0}$}%
\psfrag{Sm100}[tc][tc]{\PFGstyle $\text{-10.0}$}%
\psfrag{Sm125}[tc][tc]{\PFGstyle $\text{-12.5}$}%
\psfrag{Sm150}[tc][tc]{\PFGstyle $\text{-15.0}$}%
\psfrag{Sm25}[tc][tc]{\PFGstyle $\text{-2.5}$}%
\psfrag{Sm50}[tc][tc]{\PFGstyle $\text{-5.0}$}%
\psfrag{Sm75}[tc][tc]{\PFGstyle $\text{-7.5}$}%
\psfrag{W0}[cr][cr]{\PFGstyle $0$}%
\psfrag{Wm1}[cr][cr]{\PFGstyle $\text{-1}$}%
\psfrag{Wm2}[cr][cr]{\PFGstyle $\text{-2}$}%
\psfrag{Wm3}[cr][cr]{\PFGstyle $\text{-3}$}%
\psfrag{Wm4}[cr][cr]{\PFGstyle $\text{-4}$}%
\psfrag{Wm5}[cr][cr]{\PFGstyle $\text{-5}$}%
\psfrag{y0}[cr][cr]{\PFGstyle $0$}%
\psfrag{ym1}[cr][cr]{\PFGstyle $\text{-1}$}%
\psfrag{ym2}[cr][cr]{\PFGstyle $\text{-2}$}%
\psfrag{ym3}[cr][cr]{\PFGstyle $\text{-3}$}%
\psfrag{ym4}[cr][cr]{\PFGstyle $\text{-4}$}%
\psfrag{ym5}[cr][cr]{\PFGstyle $\text{-5}$}%